\documentclass[emulateapj]{aastex631}
\usepackage[utf8]{inputenc}
\usepackage{graphicx}
\usepackage{amsmath,amssymb}
\begin{document}
\title{Dynamical Fates of S-Type Planetary Systems in Embedded Cluster Environments}
\author{Elizabeth A. Ellithorpe}
\affiliation{HL Dodge Department of Physics \& Astronomy, University of Oklahoma, Norman, OK 73019, USA}
\author{Nathan A. Kaib}
\affiliation{HL Dodge Department of Physics \& Astronomy, University of Oklahoma, Norman, OK 73019, USA}
\affiliation{corresponding author: nathan.kaib@ou.edu}


\newpage
\begin{abstract}
The majority of binary star systems that host exoplanets will spend the first portion of their lives within a star-forming cluster that may drive dynamical evolution of the binary-planet system. We perform numerical simulations of S-type planets, with masses and orbital architecture analogous to the solar system's 4 gas giants, orbiting within the influence of a $0.5\: M_{\odot}$ binary companion. The binary-planet system is integrated simultaneously with an embedded stellar cluster environment. $\sim$10\% of our planetary systems are destabilized when perturbations from our cluster environment drive the binary periastron toward the planets. This destabilization occurs despite all of our systems being initialized with binary orbits that would allow stable planets in the absence of the cluster. The planet-planet scattering triggered in our systems typically results in the loss of lower mass planets and the excitement of the eccentricities of surviving higher mass planets. Many of our planetary systems that go unstable also lose their binary companions prior to cluster dispersal and can therefore masquerade as hosts of eccentric exoplanets that have spent their entire histories as isolated stars. The cluster-driven binary orbital evolution in our simulations can also generate planetary systems with misaligned spin-orbit angles. This is typically done as the planetary system precesses as a rigid disk under the influence of an inclined binary, and those systems with the highest spin-orbit angles should often retain their binary companion and possess multiple surviving planets. \\
\\
Keywords: stars: binaries: general, planets and satellites: dynamical evolution and stability, galaxies: star clusters: general. 

\end{abstract}

\section{Background \& Motivation}
\label{background}
\indent The dynamics of planetary systems serve as signposts of their formation and subsequent evolution. Thus the study of dynamical modeling 
of exoplanets can be thought of as analogous to archaeology: using both observations and simulations of existing stable planets, we can walk back in 
time to discover which physical processes govern the behavior of extrasolar worlds.  Dynamics offers a strong union of computational 
and observational astronomy, where an influx of exoplanet observational data has dovetailed nicely with a surge in advancements in numerical algorithms
over the last two decades. \\
\indent It has been long thought that most stars spend some of their early lives in a cluster environment, a large scale body of gas that acts as a stellar nursery
 that condenses into pre-stellar cores in the gas and dust \citep{lad03}. Recent work \citep{sad17} has also bolstered the theory \citep{kro95} that the majority of stars 
 are born in a stellar multiple, with their dense pre-stellar cores mutually bound together in a dusty envelope. 
 These multiple bound cores, likely formed
 via fragmentation \citep{bon94,bat03,tur09}, either stay together in a binary as interactions with their gaseous
 disk decrease the separation between the pair \citep{bat00,bat03}, or are pulled apart by a cluster environment \citep{kro01}. 
 While clusters can unbind so-called `soft binaries' \citep{kro01}, dynamical interactions with a cluster are not sufficient to explain the observed
 period distribution of binary stars. Thus, their nascent orbital architecture results from the fragmentation process and interactions with a circumstellar disk
 early on in their lives \citep{krobur01}, and should be considered when we discuss the formation and the dynamics of planets in a holistic manner.\\
 \indent While planets in a binary system have been 
 previously studied in depth with N-body integration, the binary that interacts with the planets is for the most part unperturbed by outside forces 
 \citep[e.g.][]{hol99, hag07},
 or, if the binary is modeled in a cluster, planets are not included \citep[e.g.][]{krobur01}. Similarly, planetary systems as they might interact with a cluster environment
 have been modeled, but often with mock fly-bys of a star rather than a fully evolved cluster \citep[e.g.][]{rec09, cat20}. A union of these two simultaneous phenomena,
 a binary's effect on planets and a cluster's effect on a binary, is necessitated by the ubiquitous nature of these interactions in young systems. With N-body simulations
 produced by hybrid symplectic integration scheme, we offer a new perspective at this complex problem. \\
\indent The presence of a binary companion has been considered in studies of planetary formation since the 1960's. Early treatments of planetary stability in a binary system worked
within the framework of the restricted 3-body problem \citep{hua60} and numerical estimates based on analytic models \citep{hep78}. Early work established that 
both the formation and the stability of planets in a binary system were possible. Later on, numerical integration to evolve these systems began in earnest, owing largely to developments in symplectic integration methods \citep[e.g.][]{wis91} that 
allowed for efficient integrations of entire planetary systems in the presence of a binary. The main question that arises in such a system is in what cases does a binary 
companion preclude the existence of stable planets. An early work that used the Wisdom-Holman scheme is the exploration of the Alpha Centauri system in \cite{wie97},
in which the authors used N-body integrations in the \cite{wis91} scheme to assess the stability of test particles in and around the Alpha Centauri triplet. \\
\indent The seminal works in the field of planetary stability in a binary system are \cite{dvo86} and \cite{hol99}. \cite{dvo86} looked at integrating systems of test
particles in 
the restricted 3-body problem with the Lie series method \citep{del85} and established an upper and lower critical orbit that began mapping a region of stability 
for a planet based on the eccentricity of the binary system. \cite{dvo86} looked at various eccentricities of the binary companion, but in all cases the mass of the binary and 
the central star were the same. 
 \citet{hol99} further built upon this stability analysis via symplectically integrating binary systems 
with both P-type (where the planet orbits around a binary star pair) and S-type planets (where planets are bound to a single star with a binary companion exterior to their orbit).
 They produced an analytic form of a planet's critical orbit based on both the eccentricity of the binary and the stellar mass ratio. Also drawing on \cite{wis91} is the work described in \cite{ina97}, in which the solar system is integrated in the presence of a 
binary. The authors exemplified the Kozai effect \citep{koz62} in which an inclined binary exchanges angular momentum with the planets, and also showed that the planets can evolve 
in concert as a rigid disk. This work revealed that the Kozai effect is a pathway that can lead to increased eccentricity of planets. 
Another analysis of the Kozai effect on planetary stability that 
takes a different, secular approach is \cite{fab07}, in which the authors integrated the motion of planets in the presence of a binary to show that tidal friction 
can circularize an orbit made eccentric by an inclined binary companion.\\
\indent The stability analysis of \citet{hol99} remains of interest to this day, with recent work such as \cite{lam18} further refining
their original critical orbit with a neural network and integrations performed in the publicly available REBOUND integrator \citep{reiliu12}. \citet{qua18} also further
examined the stability of planets in the presence of a stellar binary, focusing in particular on the planetary architectures appropriate for orbital stability 
in the $\alpha$ Centauri system.
 Analysis of planets in binaries continued with work 
in \cite{cha02} in which the authors used a mixed variable symplectic method to analyze planetary accretion both in a system where the planetary bodies orbit a primary and 
are perturbed by a distant binary and one in which they orbit around the binary system in its entirety. While \cite{cha02} utilized an integrator that allows for the presence 
of a binary, further work by \cite{beu03} introduced a symplectic scheme based on the \cite{wis91} method, called Hierarchical Jacobian Symplectic (HJS), that allows for the 
integration of a multiple system of any size (i.e., numbering more massive bodies than just a binary) as long as there is a retained
hierarchy among the masses. The presence of a distant companion has also been considered in numerical integrations of protoplanetary (debris) disks. 
\cite{rec09} used the HJS scheme to integrate the HD141596 triple star system and debris disk in the presence of stellar flybys. 
\cite{beu14} used symplectic integration to model the Fomalhaut triplet in addition to its dust belt.  

\indent On the larger scale, N-body simulations of stellar clusters have been used to study how stellar binaries interact with a cluster environment. \cite{bat03} 
used high resolution simulations of a collapsing gas cloud to study the fragmentation of gas into dense cores, 
and the subsequent evolution of these young stars into binaries. They found that very
 tight binaries are formed by the hardening of wider binaries via dynamical interactions. \cite{ada06} and \cite{pro09} used N-body simulations of moderately sized stellar clusters
  (100-10,000 members) or embedded clusters to study the impact of a cluster environment on planet formation, particularly in how protoplanetary disks may be affected by stellar raditation.
  They found that disruption of model solar systems in an evolved cluster environment 
  \citep{ada06} should be relatively rare due to the paucity of encounters between a `solar system' 
  in question and a passing star, but in younger, denser clusters the photoevaporation of disks by a cluster environment can be appreciable \citep{pro09}.\\
  \indent Also addressing the evolution of a stellar cluster is work by \cite{par09kozai} and \cite{par09binaries} in which N-body simulations are used to analyze the prevalence of 
  planets susceptible to the Kozai effect from a binary companion, and the stability of binaries in a dense cluster environment respectively. \cite{par09kozai} found that 
  around $20\%$ of all exoplanets should at one point in their lives be in the presence of a binary companion that has been sufficiently inclined by its cluster environment
  such that Kozai cycles can occur. This is an intriguing finding that 
  bolsters the idea that a binary companion perturbed by the cluster can have an appreciable affect on the planets, with the caveat that the authors only examined the evolution of binary orbits, as 
  their simulations lacked planetary bodies.
   Additionally, they particularly focused on very dense clusters similar to Orion, with a half-mass radius of only $0.1$ pc. This is not typical for embedded
  clusters, which have typical half mass radii of $\sim$0.8 pc \citep{lad03}, and therefore would have less frequent interactions between cluster stars and a particular binary.
  Nevertheless, they showed that a dense cluster environment can significantly alter the architecture of a stellar binary which in turn can affect a protoplanetary disk or mature
  planet system. \cite{par09binaries} focused on similarly dense cluster environments, but instead explored the longevity of moderately wide ($\sim$$10^3 \text{au}$) 
  to ultrawide ($> 10^4 \text{au}$) binaries. They found that cluster environments strip away all ultra-wide binary companions, and that the denser clusters, with half mass radii of 
  0.1--0.2 pc do not retain any binaries with separations $> 10^3 \text{au}$. The less dense clusters, with half mass radii 0.4--0.8 pc, do retain some of these moderately
  wide binaries. The authors noted that as ultrawide binaries are often stripped in only a few cluster crossing times, these very separated binaries may form in isolation
 An alternate channel is that ultrawide binaries are formed during cluster dissolution \citep{kou10}. \\
  \indent \cite{hao13} takes a monte carlo approach to simulating planetary systems in an
 open cluster to explore planet-planet scattering. The authors modeled stellar flybys and found that multi-planet systems are more sensitive to an open cluster 
 environment than single planet systems. In the realm of exploring planetary orbits in clusters via a model stellar fly-by is work by \cite{mal11} and \cite{bre19}.
  \cite{mal11} showed that fly-bys increase the chance of planetary ejection, while \cite{bre19} showed that a cluster
star can create a retrograde planetary orbit.  

\indent In summary, while there is an extensive body of work on the integration of planets in a binary system and the evolution of planets in a cluster environment, previous work
largely assumes that the binary is in isolation and does not evolve due to external forces; work concerning cluster environments either model interactions as a flyby 
or do not include a simultaneous binary system. The work we present here is novel in its approach in that it allows the binary companion to be altered by a cluster environment,
and that we fully integrate the passages from and evolution of the cluster rather than taking a flyby approach. 

We are largely concerned with the overall fate of binary systems and their planets during these interactions with the cluster. We find that encounters 
between the binary system and clusters stars can destabilize planets and occur at a non-negligible rate. Moreover, our
work has also revealed some intriguing changes in orbital architecture of planetary systems that have a close encounter with an excited binary companion. The remainder
of this paper is laid out into the following sections: Section \ref{methods} describes how we approach the numerical integration of these systems as well as how we set up 
the explored parameter space. Section \ref{results} details our findings from these simulations focusing on the mechanism and rate of planetary destabilizations. We then examine the aftermath of 
systems that have undergone a planetary instability in Section \ref{discuss}, highlighting potential tracers of instability and discussing the broader implications of our results. Finally, we summarize our conclusions in Section \ref{conclude}.

\section{Methods}
\label{methods}
The numerical integrator we use for this work is built on the foundation of the original \textit{Mercury} package \citep{cha99,cha02} but with changes
to allow the inclusion of multiple bound stellar mass bodies (i.e., a binary or triplet star system) and unbound stellar mass bodies (i.e. stars in a cluster environment). The details 
of the changes to the integrator are explained in \citet{kai17}.
We still use the democratic heliocentric coordinates for the planetary system, but the binary
is defined relative to the center of mass of the planetary system, and the cluster stars are defined simply with their inertial coordinates. By letting the different
bodies be treated with different coordinate systems, this version of \textit{Mercury} retains the advantages of a pure symplectic integrator while still being able 
to evolve the unbound cluster stars. That is, there is not a build up of energy error, the pseudo-Keplerian nature of the planetary system integration
is retained, and simultaneously the cluster stars and their close approaches can be integrated efficiently in a leapfrog-like scheme.  \\
\subsection{Simulation Algorithm}
In this integration scheme, we define 3 distinct types of bodies: planets 
bound to the primary, the binary companion, and cluster stars. The cluster star positions are treated with their inertial coordinates,
measured only with respect to the cluster's center of mass. The binary's position
is integrated with respect to the center of mass of the primary star system \citep{cha02}.
The planets are integrated in democratic heliocentric coordinates with respect to the primary star \citep{dun98}.
The choice of these coordinates means that the cluster stars are integrated with a purely symplectic $T+V$ leapfrog scheme. This
conserves energy and angular momentum, and is quite accurate as the timescale of cluster stars orbiting in the cluster
is much larger than the simulation time step set by the orbits of the planetary bodies. The binary and the planets 
are also integrated symplectically, although with a mixed variable symplectic Wisdom-Holman scheme \citep{wis91}. Finally, when interacting bodies
near each other and non-Keplerian terms in the Hamiltonian become
comparatively large, the bodies are integrated directly with a Bulirsch-Stoer scheme \citep{cha99}. For the planets, this occurs when passing 
within a Hill radius of one another. For cluster stars with much larger spheres of influence, the changeover to a direct integration occurs when two stars are close enough to one another that their free fall time is less than $500\times$ the simulation time-step of $100$ days. For a 1 $M_{\odot}$ star, this corresponds to a close encounter distance of $\sim$130 au.

Thus, this scheme largely preserves the angular momentum and energy conservation of a symplectic scheme while allowing
for close encounters.
Equation \ref{positions} shows how we define the position of each type of object, where $\vec{X}_A$ represents the position of the primary star,
$\vec{X}_B$ the position of the binary, $\vec{X}_{i}$ for $1\leq i \leq N_P$ is the position of a planet, and $\vec{X}_i$ for $N_P < i \leq N_P+N_S$ is the position of a cluster star.
The lowercase $\vec{x}$ represents each type of body's inertial coordinates relative to the origin/center of mass of the cluster.

\begin{equation}
    \begin{split}
        \vec{X}_A = \frac{m_a\vec{x}_A+m_B\vec{x}_B+\sum_{j=1}^{N_P}m_j\vec{x}_j}{m_A+m_B+\sum_{j=1}^{N_P}m_j} \\
        \vec{X}_i = \vec{x}_i-\vec{x}_A \text{   for   } 1\leq i \leq N_P \\ 
        \vec{X}_B = \vec{x}_B-\frac{m_A\vec{x}_A+\sum_{j=1}^{N_P}m_j\vec{x}_j}{m_A+\sum_{j=1}^{N_P}m_j} \\
        \vec{X}_i = \vec{x}_i \text{   for   } N_P < i \leq N_P+N_S
     \end{split}
\label{positions}
\end{equation}
As such, we define some useful position vectors for changing between reference frames. $\vec{s}$ marks the barycenter of the planets
relative to the primary host star and $\vec{\Delta}$ marks the inertial position of the primary star as defined with the new coordinates ($\vec{X}$'s).  \\
\begin{equation}
\begin{split}
    \vec{s} = \frac{\sum_{i-1}^{N_P}m_i\vec{X}_i}{m_A+\sum_{i-1}^{N_P}m_i}\\
    \vec{\Delta} = \vec{X}_A-\frac{\sum_{i=1}^{N_P}m_i\vec{X}_i+m_B(\vec{X}_B+\vec{s})}{m_A+m_B+\sum_{i=1}^{N_P}m_i}
\end{split}
\end{equation}
The corresponding conjugate momenta for the bodies is as follows, again where the lowercase terms are for the inertial case and the uppercase terms
are the momenta for the new coordinates:

\begin{equation}
    \begin{split}
        \vec{P}_A = \vec{p}_A+\vec{p}_B + \sum_{j=1}^{N_P}\vec{p}_j \\
        \vec{P}_i = \vec{p}_i-m_i\frac{\vec{p}_A+\sum_{j=1}^{N_P}\vec{p}_j}{m_A+\sum_{j=1}^{N_P}m_j} \text{   for   } 1\leq i \leq N_P \\
        \vec{P}_B = \vec{p}_B -m_B\frac{\vec{p}_A+\vec{p}_B+\sum_{j=1}^{N_P}\vec{p}_j}{m_A+m_B+\sum_{j=1}^{N_P}m_j} \\
        \vec{P}_i = \vec{p}_i \text{   for   } N_P < i \leq N_P+N_S
    \end{split}
\end{equation}
\indent In this symplectic scheme, we must consider how Hamilton's equations evolve for all members of the system. We consider mutual gravitation
between the primary star, binary star, planets and cluster stars. We also consider the gravitational tide from the gas of the 
Plummer potential \citep{plu11} that we assume to be binding the cluster stars together. We include the derivation for the Plummer potential here. We treat 
the force from a Plummer distribution as a constant background potential whose origin is the center of mass of the cluster stars. 
As the effect of the Plummer potential is dependent on the inertial position of each massive body within the cluster, the 
calculation in our chosen (non-inertial) coordinates is fairly complicated. The expression for the force from the sphere of gas is different for each type of body,
namely the planets, the binary, and the cluster stars. We begin with the expression for the mass distribution of a Plummer sphere in inertial coordinates, where
$\vec{x}$ is defined relative to the cluster's center-of-mass. 
\begin{equation}
\rho (\vec{x}) = \frac{3M_0}{4\pi a_0} \times \biggl(1+\frac{|\vec{x}|^2}{a_0^2} \biggr)^{-5/2}
\end{equation}
This mass distribution, where $M_0$ is the total mass in gas of the Plummer sphere and $a_0$ is a scale parameter that sets the size of the flat core region results
in a gravitational potential of:
\begin{equation}
    \Phi(\vec{x}) = -\frac{GM_0}{(|\vec{x}|^2+a_0^2)^{1/2}}
\end{equation} 
The potential energy of our system can then be written as:
\begin{equation}
    \begin{split}
        V_{Plum} = -\sum_{i=N_p+1}^{N_p+N_s}\frac{GM_{0}m_i}{(|{\vec{X}_i}|^2+a_0^2)^{1/2}} - 
        \sum_{i=1}^{N_p}\frac{GM_{0}m_i}{(|(\vec{\Delta}+ \vec{X_i})|^2+a_0^2)^{1/2}} \\ 
        -\frac{GM_{0}m_B}{(|(\vec{\Delta} + \vec{s} + \vec{X}_B)|^2+a_0^2)^{1/2}} 
        - \frac{GM_{0}m_A}{(|\vec{\Delta}|^2+a_0^2)^{1/2}} 
    \end{split}
\end{equation}
For clarification, the inertial position with which the strength of the Plummer potential is measured in simulation coordinates is $\vec{X}_i$ for the cluster stars,
$\vec{\Delta}+\vec{X}_i$ for the planets, $\vec{\Delta}+\vec{s}+\vec{X}_B$ for the binary, and $\vec{\Delta}$ for the primary.
We can then calculate the acceleration due to the Plummer sphere on each body (indexed with $i$) for each cartesian coordinate (indexed with $u$) via:
\begin{equation}
    m_i\frac{dv_{i,u}}{dt} = -\frac{\partial V_{plum}}{\partial x_{i,u}}
\end{equation}
\subsubsection{Cluster Stars}
The positions of the cluster stars in the integration coordinates are exactly their inertial coordinates in this integration scheme 
(i.e., they are measured relative to the origin). This makes finding the acceleration due to the Plummer tide much more straightforward 
than for the planets and the binary star pair:
\begin{equation}
    \vec{X}_i = \vec{x}_i \text{  for  } i>N_p
\end{equation}

\begin{equation}
    \frac{dv_{i,u}}{dt} = 
    -\frac{GM_{0}X_{i,u}}{(|\vec{X_i}|^2+a_0^2)^{3/2}}
\end{equation}
We can see that the acceleration due to the Plummer sphere on the cluster stars is simply a direct term based on the mass of the gas 
enclosed by the cluster star's current position.
\subsubsection{Planets}
Recall that $\vec{\Delta}$ is a function of the position of the planets. 
\begin{equation}
    \frac{\partial \vec{\Delta}}{\partial X_i} = -m_i\frac{1}{m_A+\sum_{k=1}^{N_P}m_k}
\end{equation}
This dependence on the planets' positions leads to a more complicated expression for the Plummer potential on the planetary bodies.


\begin{equation}
    \begin{split}
    \frac{dv_{i,u}}{dt} = 
    \frac{GM_{0}}{m_A+\sum_{j=1}^{N_P}m_j}\biggl[\sum_{k=1}^{N_P}\frac{m_k(\Delta_u+X_{k,u})}{(|\vec{\Delta}+\vec{X_k}|^2+a_0^2)^{3/2}}+
    \frac{m_A\Delta_u}{(|\vec{\Delta}|^2+a_0^2)^{3/2}} \biggr]\\ - \frac{GM_{0}(\Delta_u+X_{i,u})}{(|\vec{\Delta}+\vec{X_i}|^2+a_0^2)^{3/2}}   
    \end{split}
\end{equation}

\subsubsection{Binary Companion}
The vector $\vec{\Delta}$ is also a function of the position of the binary: 
\begin{equation}
    \frac{\partial \vec{\Delta}}{\partial X_B} = \frac{-m_B}{m_A+m_B+\sum_{i=1}^{N_P}m_i}
\end{equation}
Thus, as with the planets, the expression for the acceleration of the binary star due to the Plummer potential is more complicated than what we showed previously for the cluster stars:


\begin{equation}
    \begin{split}
        \frac{dv_{B,u}}{dt} = \frac{GM_{0}}{m_A+m_B+\sum_{j=1}^{N_P}m_j}\biggl[\sum_{k=1}^{N_P}\frac{m_k(\Delta_u+X_{k,u})}{(|\vec{\Delta}+\vec{X}_k)|^2+a_0^2)^{3/2}}\\ 
        + \frac{X_{B,u}+s_u+\Delta_u}{(|\vec{\Delta}+\vec{X}_B+\vec{s}|^2+a_0^2)^{3/2}} +\frac{m_A\Delta_u}{(|\vec{\Delta}|^2+a_0^2)^{3/2}}\biggr] - 
        \frac{GM_{0}(X_{B,u}+s_u+\Delta_u)}{(|\vec{\Delta}+\vec{X}_B+\vec{s}|^2+a_0^2)^{3/2}}
    \end{split}
\end{equation}

This modified version of the \textit{Mercury} simulation package is publicly available for download at \url{https://github.com/nathankaib/Mercury_StarCluster}. 

\subsection{Cluster Environments}
First, we must decide how to build our stellar clusters. Our clusters consist of many discrete stellar mass bodies embedded within a more massive background Plummer potential to mimic the gravitational effects of a dominant gaseous component to the clusters. To design the parameters of the stellar cluster, we turn to the embedded cluster catalog compiled by \citet{lad03}, which \citet{ada06} find is mostly comprised of clusters with 100--1000 stars.

While 1000-star clusters begin to have prohibitive runtimes, we employ small, medium, and large cluster environments with 122, 221, and 509 stars, respectively. These have total stellar masses of 80, 160, and 341 M$_{\odot}$. In embedded clusters, the stellar mass only represents a small fraction of the total cluster mass, as \citet{lad03} estimate embedded cluster star formation efficiencies between $\sim$10--25\%. To bracket the low end of this range, our small and medium clusters have Plummer masses of 1000 and 2000 M$_{\odot}$, yielding star formation efficiencies of 7.4\%. Meanwhile, our large cluster employs a Plummer mass of 1364 M$_{\odot}$, yielding a higher star formation efficiency of 20\%. To highlight the role of the non-stellar potential, we also include a version of our medium cluster with all stellar masses reduced by an order of magnitude, implying a star formation efficiency of just 0.74\%.

To set the Plummer radii of our potentials (and stellar distributions), we again consult \citet{lad03}. Using this catalog, \citet{ada06} estimate that 122-, 221-, and 509-star clusters (our small, medium, and large clusters) have typical radii of 0.63--1.1, 0.85--1.5, and 1.3--2.3 pc, respectively. For our small and medium cluster environments, we set our Plummer radius to 0.62 pc, implying a half-mass radius of $\sim$0.8 pc. For the large cluster environment, we set the Plummer radius to 1.38 pc, yielding a half-mass radius of 1.8 pc.

Our cluster environment parameters are summarized in Table \ref{tab:clusters}. The goal of this work is to probe a significant swath of cluster parameter space relevant to many planetary systems and not necessarily to the solar system. However, it may be instructive to compare our parameters with constraints on the solar birth cluster, since our work assumes planetary system architectures like our own and it has been speculated that the early Sun may have possessed a binary companion \citep{sirajloeb20}. Primarily relying on cluster perturbations to generate the orbit of Sedna while preserving the stability of the giant planets, \citet{adams10} concludes that the Sun's birth cluster contained 1500--7100 stars, or roughly an order of magnitude more than our clusters. However, this assumes a cluster residence time of $\gtrsim$10 Myrs. Using the Kuiper belt inclination induced by flybys of cluster stars, \citet{bat20} find that the product of the solar birth cluster residency and stellar density must be below $\sim$$2\times10^4$ Myr/pc$^3$. The values for our clusters are 1--2 orders of magnitude less than this upper limit, so it again appears that much of the allowable parameter space for the solar birth cluster is more populous or denser than our modeled environments.

\begin{table}
\centering
\begin{tabular}{c c c c c c c}
\hline
Cluster & Plummer & Stellar & Star Formation & Plummer & N$_{*}$ & $\bar{n_{*}}$ \\
 & Mass & Mass & Efficiency & Radius & & \\
 & (M$_{\odot}$) & (M$_{\odot}$) & (\%) & (pc) &  & (pc$^{-3}$) \\
\hline
Small & 1000 & 80 & 7.4 & 0.62 & 122 & 47.4 \\
Medium & 2000 & 160 & 7.4 & 0.62 & 221 & 86.0\\
Large & 1364 & 341 & 20 & 1.38 & 541 & 22.2\\
NonStellar & 2000 & 16 & 0.74 & 0.62 & 221 & 86.0\\
\hline
\end{tabular}
\caption{Parameter summary of simulated cluster environments. From left to right, columns are: environment name, mass of Plummer potential, mass of stellar component, implied star formation efficiency, number of stars, mean volume density of stars inside half-mass radius.}
\label{tab:clusters}
\end{table}

\subsection{Simulation Setup and Progression}

With cluster parameters specified, our cluster assembly and integration prescription is largely based on the methods outlined in \citet{lev10}, with some minor differences, as noted below. The masses of the stars are sampled from the \citet{cha03} piecewise IMF, where we set an upper mass limit of $150\: M_{\odot}$ \citep{wei04}. Then, the virial speeds of each star are calculated based on their position and we assign them a random initial velocity vector at $8\%$ of the virial speed \citep{lev10}. Once the cluster bodies are positioned following the cluster Plummer profile, we generate different simulations, where for each simulation a random cluster star is converted into the role of the primary, such that the starting position of the planetary/binary system within the cluster is randomized.

A set of 4 gas giants, analogous to the orbits and masses of Jupiter, Saturn, Uranus, and Neptune, are set around a 1 M$_{\odot}$ primary star, and a coplanar 0.5 M$_{\odot}$ binary companion is added as well. We choose to fix our binary mass at this fiducial subsolar value to limit the number of parameters being simultaneously varied. Higher companion masses will likely enhance the destabilizing power of the binary companion, while lower ones should diminish it. 

Before integration within a cluster environment, all of our planet+binary systems are first verified via numerical integration to be inherently stable for $10$ Myr in isolation. If they are not stable, the initial conditions are discarded. If they do prove to be stable, they are then integrated for 10 Myrs within the cluster environment in concert with all of the other cluster stars. After 5 Myrs of this 10-Myr integration, the Plummer potential is instantaneously dispersed, leaving only the stars and planets. This corresponds to the physical dispersal of the cluster gas, and past modeling works have employed similar dispersal timescales \citep{ada06, pro09, lev10, bras12}.
 
\subsection{Binary Orbits}

\begin{table}
\centering
\begin{tabular}{c c c c c}
\hline
Name & Cluster & Binary & Binary & Number of\\
 & Environment & Eccentricities & Semimajor Axes & Simulations \\
 &  &  & (au) & \\
\hline
Small\_Fixed & Small & $e=0.5$ & 300--800 & 600\\
Medium\_Fixed & Medium & $e=0.5$ & 300--800 & 600\\
Large\_Fixed & Large & $e=0.5$ & 300--800 & 600\\
Medium\_Cold & Medium & $f(e)\propto \sqrt{e}$ & 300--800 & 600\\
Medium\_Uniform & Medium & $f(e)\propto e$ & 300--800 & 600\\
Medium\_Thermal & Medium & $f(e)\propto e^2$ & 300--800 & 600\\
Medium\_Wider & Medium & $f(e)\propto e^2$ & 900--1300 & 500\\
NonStellar\_Fixed & NonStellar & $e=0.5$ & 300--800 & 600\\
NonStellar\_Wider & NonStellar & $f(e)\propto e^2$ & 900--1300 & 500\\
\hline
\end{tabular}
\caption{Summary of simulation sets. From left to right, columns are: simulation set name, cluster environment of simulation set (see Table \ref{tab:clusters}), binary eccentricity distribution, and range of binary semimajor axes (spanned in 100 au intervals), and total number of simulations.}
\label{tab:simsum}
\end{table}

The influence of the binary on our planetary systems and its susceptibility to perturbation from our cluster environments of course strongly depends on its orbit. Our choices for binary orbits vary across our different sets of simulations. To initially limit the number of parameters being varied within a simulation set, we first perform three sets of 600 simulations in which the binary eccentricity is fixed at 0.5. Meanwhile, in each simulation set, the binary semimajor axes are varied in increments of 100 au between 300 and 800 au, with 100 different simulations (different starting cluster positions) performed at each binary semimajor axis. (For semimajor axes closer than $300$ au the planets are typically immediately unstable, and some binaries beyond $800$ au begin to dissociate before cluster gas dispersal.) These fixed eccentricity runs allow us to better isolate the importance of binary semimajor axis and cluster environment.

An additional three sets of simulations explicitly study the importance of the binary's initial eccentricity within the medium cluster environment. The three different sets employ different distributions of binary eccentricity: uniform in $e$, uniform in $e^2$ \citep[a thermalized distribution;][]{amb37}, and uniform in $e^{1/2}$. For each individual simulation, we again first verify that the choice of binary eccentricity between $0$ and $1$ from the requisite distribution resulted in a system of stable planets for $10$ Myr without the presence of the cluster environment. This gave us an approximate upper bound of $e=0.8$ for the widest 800 au binaries and $e=0.6$ for the tightest 300 au binaries. Especially in the case of simulations where the planets are on the edge of stability, we operate under the assumption that planets can form in the presence of such an eccentric binary companion. This can be physically explained either that we start the $10$ Myr simulation at a snapshot where the cluster has perturbed the binary to its high starting eccentricity and we are examining how long the planets remain stable after further perturbations, or that when the planets are forming in a gas disk, the eccentricity perturbations from cluster stars are damped out by the disk material. Either way, we are embedding a system that is stable in isolation, and examining how the cluster affects the binary from that point.

We also run another follow-on set of simulations that explore slightly larger binary semimajor axes. As mentioned earlier, the binary dissociation times decrease with binary semimajor axes. To destabilize the planets from an $e=0.5$ starting eccentricity, the binary periastron has to decrease dramatically in a widely separated binary. However, our thermalized runs begin many binaries on higher eccentricities nearer to the point of planetary instability. To study whether more weakly bound binaries could also trigger a planetary system instability, this set of simulations spans binary semimajor axes between 900 and 1300 au in semimajor axis increments of 100 au. These binaries are also immersed in the medium cluster environment described above.

Finally, we perform two more sets of simulations in the reduced stellar mass version of our medium cluster environment. These are meant to explore the importance of the Plummer gas potential to our systems' dynamics. The first set employs binaries with a fixed eccentricity of 0.5 and semimajor axes from 300--800 au, and the second set employs a thermal binary eccentricity distribution with semimajor axes from 900--1300 au. Our various sets of simulations are summarized in Table \ref{tab:simsum}.

\section{Results}
\label{results}

\begin{table}
\centering
\begin{tabular}{c c c}
\hline
Name & Planetary & Binary\\
 & Disruption Rate & Dissociation Rate\\
 &  (\%) & (\%) \\
\hline
Small\_Fixed & 5.5 & 8.2\\
Medium\_Fixed & 9.0 & 13.0\\
Large\_Fixed & 7.3 & 8.8\\
Medium\_Cold & 6.7 & 11.5\\
Medium\_Uniform & 10.8 & 12.7\\
Medium\_Thermal & 14.5 & 13.2\\
Medium\_Wider & 14.4 & 22.8\\
NonStellar\_Fixed & 2.0 & 0.3\\
NonStellar\_Wider & 8.0 & 0.8\\
\hline
\end{tabular}
\caption{Summary of instability rates in our simulation sets. From left to right, columns are: simulation set name, percentage of destabilized planetary systems, and percentage of unbound binary companions.}
\label{tab:instabsum}
\end{table}

\begin{figure}
    \centering
    \includegraphics[width=0.5\textwidth]{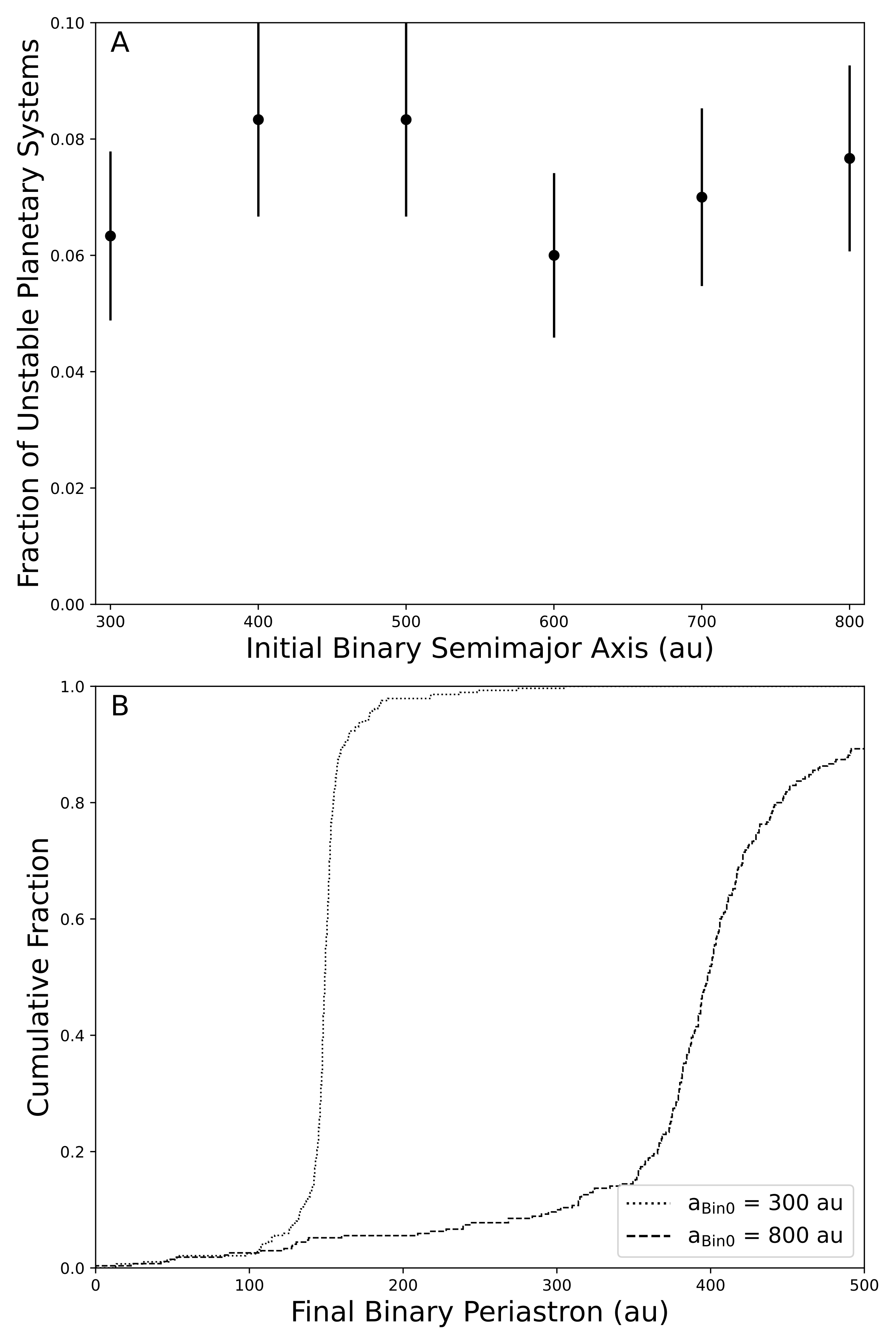}
    \caption{\textbf{A:} The fraction of planetary systems that experienced one or more planetary ejections as a function of initial binary semimajor axis with fixed initial binary eccentricity of 0.5. Different cluster mass simulations are co-added and error bars are 1$\sigma$ Poisson uncertainties. \textbf{B:} The cumulative distribution of final binary periastra for binaries beginning with $e=0.5$. Distributions are obtained via co-adding all three different cluster mass simulations.}
    \label{fig:BinSMA}
\end{figure}

The rates that planetary systems are destabilized are shown in Table \ref{tab:instabsum}. We see no strong dependence on the cluster masses we employ. Amongst our fixed eccentricity binaries, we find that 5.5\%, 9.0\%, and 7.3\% of the systems in our small, medium, and large clusters experience planetary system instabilities (lose at least one planet via collision or ejection). Surprisingly, we also do not observe a strong dependence on initial binary semimajor axis. In Figure \ref{fig:BinSMA}A, we plot the fraction of planetary systems that undergo an instability as a function of binary semimajor axis for our fixed binary eccentricity runs. As can be seen, there is no clear trend across binary semimajor axes of 300--800 au, with instability rates hovering between $\sim$5--9\%. We can see why this is the case in Figure \ref{fig:BinSMA}B, where we examine the final periastron distributions for $a=300$ au binaries and $a=800$ au binaries. Here we see that the two distributions are radically different, retaining signatures of their identical initial eccentricities but differing semimajor axes. However, at the low periastron tail, the two samples have nearly identical fractions of systems with periastra below $\sim$100 au. Coincidentally, periastron passages within 100 au are what are typically capable of destabilizing our planetary systems, where the instability probability quickly falls off for more distant periastron passages. Thus, the greater periastron evolution experienced by our $a=800$ au binaries roughly offsets the effects of the lower initial periastra of our $a=300$ au binaries.

\begin{figure}
    \centering
    \includegraphics[width=0.65\textwidth]{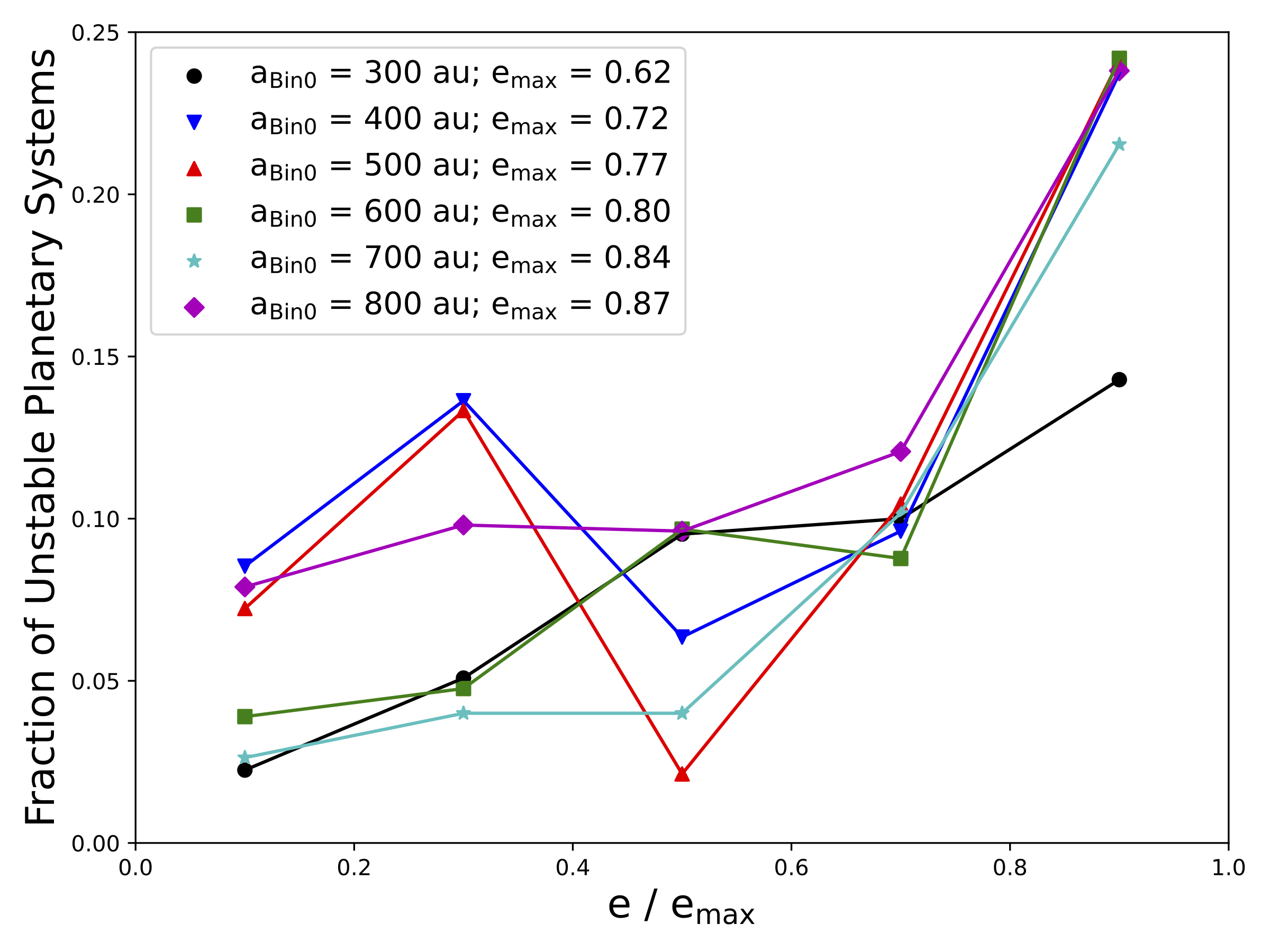}
    \caption{Planetary instability rate as a function of initial binary eccentricity (normalized by the maximum initial eccentricity employed for a given binary semimajor axis) in our Medium\_Cold, Medium\_Uniform, and Medium\_Thermal runs. Different lines and symbols correspond to different binary semimajor axes, as detailed in the legend. Legend also lists the maximum initial binary eccentricity for each semimajor axis.}
    \label{fig:binecc_instab}
\end{figure}

For our variable eccentricity simulations, we see a clearer sign that instability rates vary with different initial distributions of binary eccentricity. The fraction of planetary instability is $6.7\%$ for the low eccentricity distributions, $10.8\%$ for the uniform case, and $14.5\%$ for the thermalized case, steadily increasing as the initial binary eccentricities are biased toward higher values. We can see this dependence more obviously in Figure \ref{fig:binecc_instab}. Here we co-add the results of our Medium\_Cold, Medium\_Uniform, and Medium\_Thermal runs and then separate them by binary semimajor axis. The fraction of planetary systems that have undergone instability is then plotted against the initial binary eccentricity normalized by the maximum binary eccentricity used for a given binary semimajor axis. (Since we require systems to be stable in isolation, different binary semimajor axes allow for different maximum eccentricities.) This plot demonstrates that the probability of planetary instability increases by a factor of $\sim$3--5 across the range of eccentricities employed for each binary semimajor axis.

Although we see a clear dependence of planetary instability probability on the binary's eccentricity, we again do not see a strong dependence on binary semimajor axis. Even though some of our wider binaries begin to be dissociated by the end of the cluster's life and also need to attain a more extreme eccentricity to disrupt their planets, this appears to be roughly offset by the greater orbital variation these binaries incur due to cluster perturbations. This raises the possibility that significant planetary instability rates persist to wider binaries. Our Medium\_Wider simulations suggest that this is the case. Extending the binary semimajor axes from 800 au to 1300 au (using a thermal binary eccentricity distribution), we see no sign of a significant drop off in planetary instability rates, with 14.4\% of systems undergoing an instability. This is in spite of the fact that our widest binaries ($a=1300$ au) need to typically reach a much more extreme eccentricity of $\sim$0.93 to destabilize the planets compared to our tightest binaries, which only need to typically exceed $e=0.62$.

\begin{figure}
    \plotone{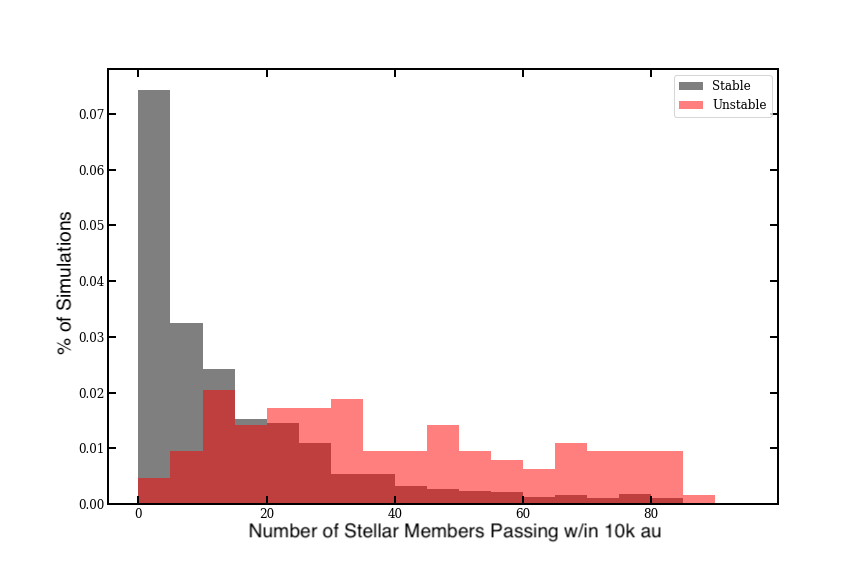}
    \caption{A normalized histogram containing results from all 3 suites of simulation depicting the distribution of number of `close' (within 10k au) 
    encounters the planetary system has with cluster stars. The systems with unstable planets are in red, showing a tail of multiple close encounters. The 
    stable systems are shown in black.}
    \label{fig:close_hist}
\end{figure}

A uniting factor that we can point to in the case of destabilized systems is the number of close encounters a system has with cluster stars, as the perturbing of the binary by the stochastic motion of the cluster can produce planetary instability. This is what we see in the vast majority of cases: interactions with passing cluster stars drive the binary companion to a very eccentric orbit, allowing for close pericenter passages with the primary and its planets. \cite{kaib13} found a similar result in an exploration of wider binaries in the galactic field wherein even distant binary companions can be perturbed to eccentric orbits with close pericenter approaches to an inner planet system.
In fact, in our simulations the frequency of close stellar encounters seems to be the most important factor in whether or not a system is destabilized. Unstable systems have a clear preference for multiple, close interactions with cluster stars. This is shown in the histogram in Figure \ref{fig:close_hist}, where we count the number of times cluster stars pass within $10^4$ au of our binary system. A clear takeaway from this figure is that destabilized systems typically experience far more close encounters with stellar cluster members than stable systems. The probability of having a large number of close stellar encounters within the cluster is largely dependent on the positional history of our systems within the cluster environment. For each of our simulations, we measure the distance from the primary to the cluster center-of-mass at each time output. For systems that lose at least one planet, this distance's median value is over 4 times smaller compared to those that remain stable.

\begin{figure}
\centering
    \includegraphics[scale=0.6]{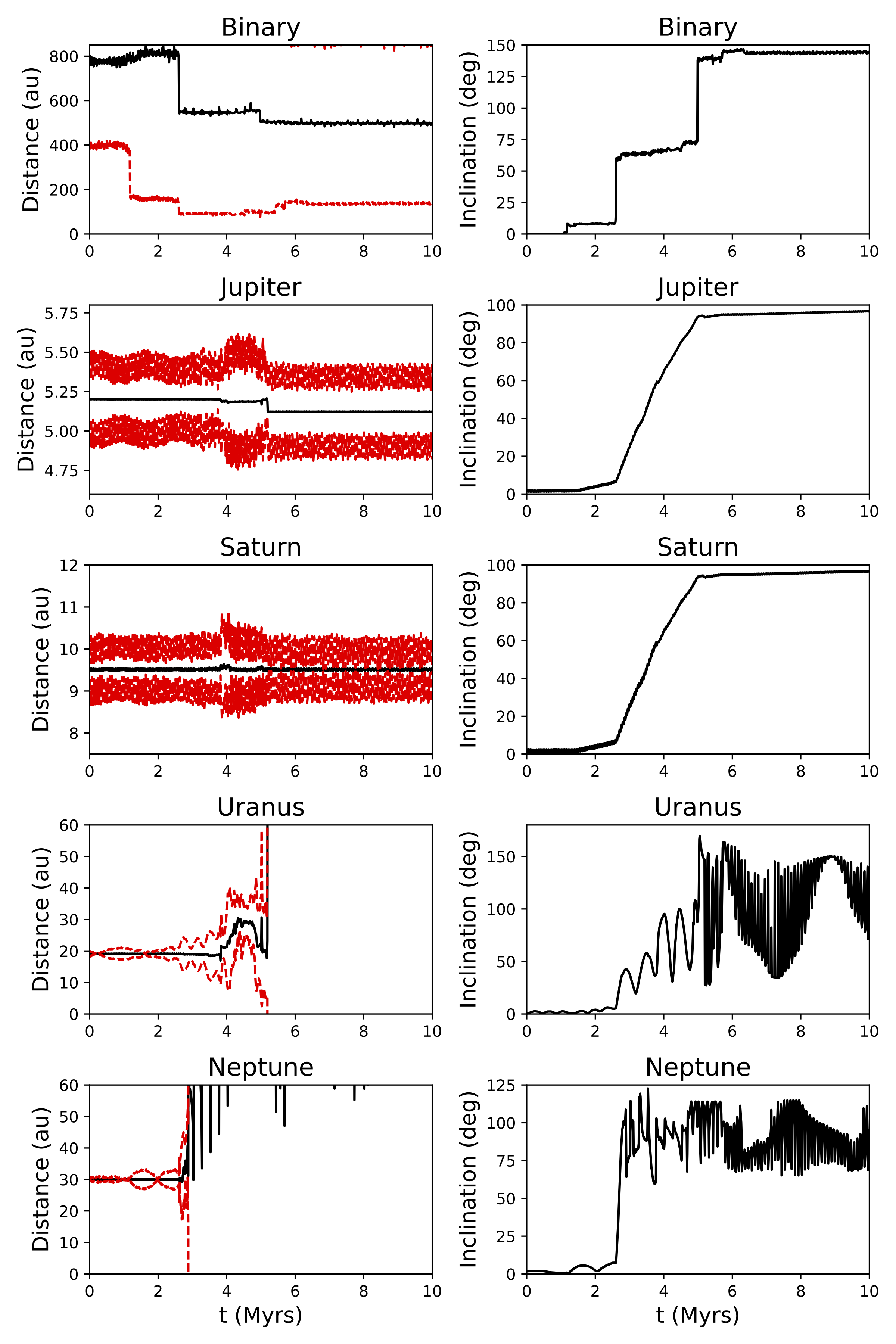}
    \caption{The orbital evolution of the 4 giant planets and binary companion. The left panels are the distance of the object from the primary,
    where the solid black line is the semi-major axis and the dashed red lines are the pericenter/apocenter of the body. This particular system results in a very high
    predicted spin-orbital angle for the Jupiter and Saturn mass bodies. Note the multiple `jumps' as the binary companion 
    enters a close pericenter approach. These jumps are from encounters with cluster stars in which angular momentum is exchanged. In this
    simulation both the Uranus and Neptune mass planets become unbound and the binary companion remains bound.}
    \label{fig:stepbinary}
\end{figure}

Figure \ref{fig:stepbinary} illustrates why a large number of encounters with cluster stars enhances the probability of a planetary instability. The example system shown here retains its binary companion, but it evolves from an initial semi-major axis of $a=800$ au to $a\approx 500$ au. At the same time, the binary periastron evolves from 400 au to under 100 au, leading to strong perturbation of the planetary orbits. During this inwards binary evolution, Jupiter and Saturn are pulled to $\sim$$100^{\circ}$ inclination (as the binary's orbital inclination also increases to over 130$^{\circ}$), while Uranus and Neptune are ejected from the system. This particular simulation is a good representative of many common characteristics we observe in unstable systems: 1) the binary star moves inward, and its close periastron approaches drive the loss of the planets. 2) The massive inner planets are the most likely survivors of this event. 3) The remaining bound planets are on altered orbits following the binary's periastron approach, having both higher inclination and eccentricity on average than their stable counterparts. As frequent close encounters from cluster stars are the main driver in these orbital changes, the architecture of planetary system is largely `frozen in' once the cluster becomes unbound after $5$ Myr \citep{all07}, as the frequency of stellar encounters drops precipitously. A similar lack of stellar encounters when the stars are unbound from the gas was seen in \cite{ada06}. 

In $73\%$ of our destabilized systems, at least one planet is left behind, which is good news for observational signals of these disruption events. This may allow us to make inferences about orbital changes due to binary destabilization processes in surviving exoplanets. Figure \ref{fig:stepbinary} suggests that planets surviving these binary-triggered instabilities may reside on exceptionally eccentric and/or inclined orbits. 

We can also examine the final fate of the binary companion, and we use our fixed binary eccentricity runs to do so. For unstable systems with higher planetary inclinations, this effect is often tied to a similar change in the binary orbit. Among our unstable systems that retain a binary companion and at least one planet at the end of the simulation, the inner planet has a median inclination of 38.4$^{\circ}$. If we just look at the systems with planetary inclinations higher than this median, we find that the final median binary inclination is 52.1$^{\circ}$. However, when we look at unstable systems in the bottom half of planetary inclinations, we find a much lower final median binary inclination of just 12.1$^{\circ}$. 

Of course, not all binary companions remain bound until the end of our simulations. Out of our 1800 fixed eccentricity simulations, binary disruption is relatively rare, with 10\% of binaries becoming dissociated. However, amongst systems that underwent an instability, the rate of binary dissociation is much higher at 56\%. (This is due to the fact that exceptional numbers of close encounters with cluster stars drive the binary evolution required to trigger a planetary instability.) Thus, for the majority of unstable systems, the binary that perturbed the planets is eventually lost to the cluster. Many single-star exoplanet systems should exist in which the planetary orbital architecture exhibits hallmarks of a violent instability, yet the destabilizing binary companion responsible for this is no longer associated with the system.

\indent It is becoming recognized that the exoplanets with the highest known eccentricities have a higher probability of possessing a stellar companion \citep{ven21}. In fact, $\sim$50\% of exoplanets whose eccentricities exceed 0.8 reside in multi-star systems. Our instabilities triggered by a quickly vacated binary companion may also explain a substantial fraction of the $\sim$50\% of such eccentric exoplanets that are not observed to currently reside in a stellar multiple.


\section{Discussion}
\label{discuss}
\subsection{Orbital Architecture Changes}
\label{orbchange}
 The orbital architectures of surviving planets in unstable (lose at least one planet) versus stable (retain all four planets) have at least two potentially observable signposts: the eccentricity and spin-orbit angle of the surviving planets. In the following analysis, we take the final planetary orbital inclinations as a proxy for their spin-orbit angles. This makes two implicit assumptions. The first is that the planets are formed on orbits whose angular momentum is aligned with the stellar spin axis. The second is that tidal forces do not subsequently realign the spin axes of our stars with the orbital angular momentum vectors of our planets. Although the sample of known Hot Jupiters with large spin-orbit angles suggests that tidal realignment can occur after spin-orbit angle excitation, the timescale for realignment is not well-constrained, and such timescales would be much longer in our simulated systems whose planets orbit at multiple au \citep{win10, hamerschlaufman22}.
 
\begin{figure}[h!]
    \centering
    \includegraphics[scale=0.5]{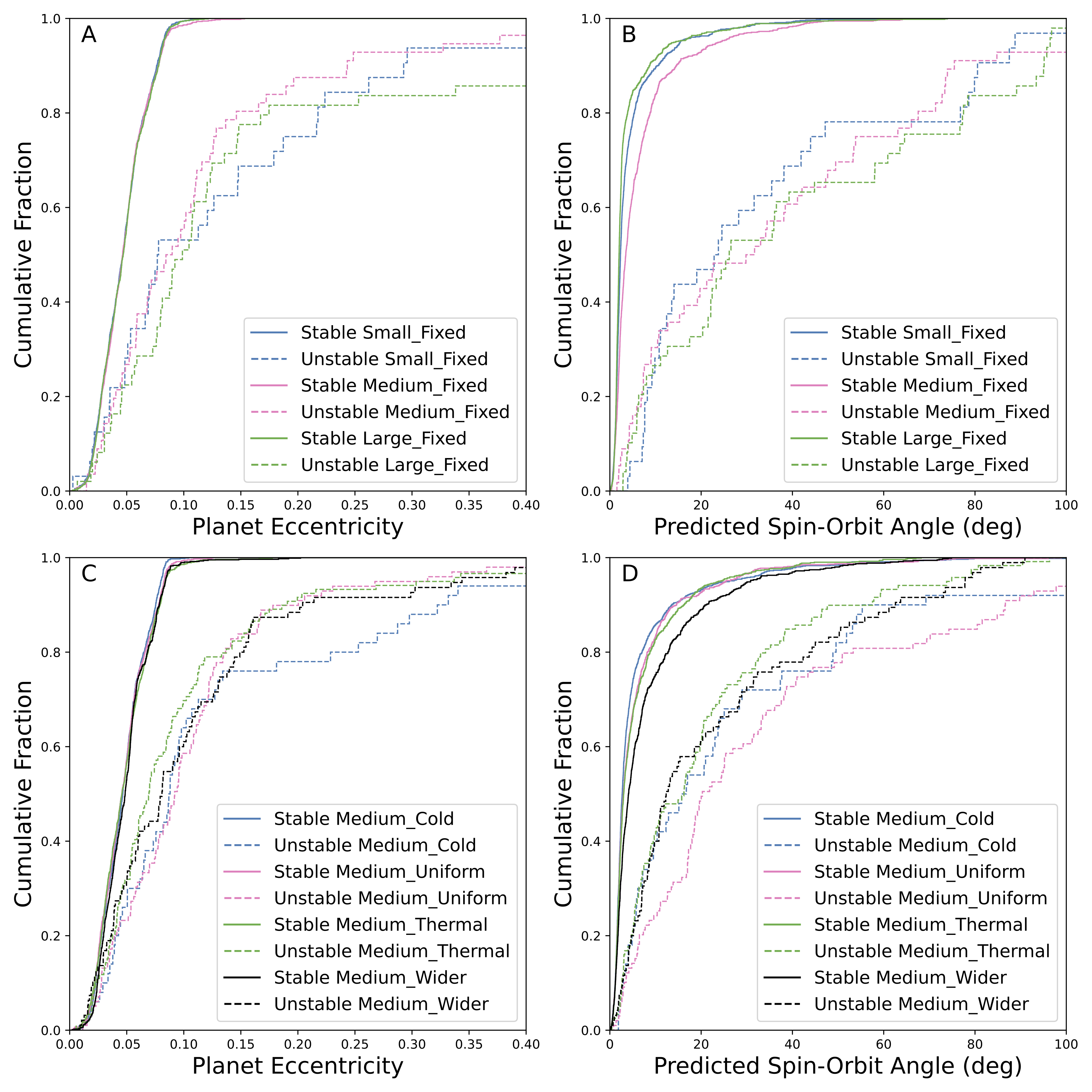}
    \caption{{\bf A:} Cumulative distributions of eccentricities of Jupiter and Saturn analogs in stable (solid lines) and unstable (dashed lines) systems from our Small\_Fixed, Medium\_Fixed, and Large\_Fixed simulations. {\bf B:} Cumulative distributions of the predicted spin-orbit angles of the planets from Panel A. {\bf C:} Cumulative distributions of eccentricities of Jupiter and Saturn analogs in stable (solid lines) and unstable (dashed lines) systems from our Medium\_Cold, Medium\_Uniform, Medium\_Thermal, and Medium\_Wider simulations. {\bf D:} Cumulative distributions of the predicted spin-orbit angles of the planets from Panel C. }
    \label{fig:160_ecc_inc}
\end{figure}

The distributions of eccentricity and the predicted spin-orbit angle of the inner gas giants in unstable systems are statistically distinct from their stable counterparts. High eccentricity planets, like the ones we see in our unstable systems, are thought to have a dynamical source. First posited in \citet{ras96}, planet-planet scattering interactions can lead to ejections of some planets and a higher eccentricity for planets that remain.
  In our systems, often these surviving planets are the two inner gas giants. The Jupiter and Saturn mass planets 
  thus act as good markers of a past instability, and our analysis will focus on their final orbits in each simulation.
The distributions of the final eccentricity and inclination of the inner giants are shown in Figure \ref{fig:160_ecc_inc}. Each of our batches of simulations tells a similar story. The median eccentricities of Jupiter and Saturn analogs in our stable systems are around 0.045, but in systems that have undergone binary-triggered instabilities this median value grows by 50--100\%. In addition, a small fraction of unstable systems are driven to more exceptional eccentricities. Depending on the particular simulation batch, 1--3\% percent of our surviving gas giants achieve eccentricities above 0.5.

We see similar trends in the evolution of planetary inclinations and assumed spin-orbit angle. In stable systems, the Jupiter and Saturn analogs have a median predicted spin-orbit angle of 2--4$^{\circ}$. However, in instability survivors this median value increases to 15--30$^{\circ}$, depending on the particular simulation batch. Often the two inner planets will undergo a coupled evolution as if in a rigid disk (shown in work by \cite{ina97}), reaching similar inclinations after the instability. In addition, approximately 5\% of our surviving planets in unstable systems attain spin-orbit angles in excess of 90$^{\circ}$.

Interactions with the binary companion, whether it remains bound to the primary or not, destabilize the planets via an exchange of orbital angular momentum. If the planets stay bound, this interaction tends to heat up their orbits to higher eccentricities and inclinations. However, when we examine our unstable systems, we find that final planetary inclinations and eccentricities are not particularly strongly correlated ($r<0.1$). This is because the most common instability leads to the ejection of one or both ice giants, while the Jupiter and Saturn analogs survive. The scattering of ice giants does not necessarily strongly drive the excitation of Jupiter and Saturn analog eccentricities. On the other hand, the binary evolution necessary to trigger ice giant ejections also often elicits significant rigid disk precession of the Jupiter and Saturn analogs, significantly increasing the spin-orbit angle \citep{kai11}. If we instead limit ourselves to systems that only have 1 surviving planet (nearly always Jupiter after a Saturn ejection), we find a stronger correlation ($r=0.37$) between final inclination and eccentricity. In addition, the median planetary eccentricity is much higher at 0.25, and this sample also contains 2/3 of our planets that finished with eccentricities above 0.5. 

\subsection{A Significantly Misaligned Population}
\begin{figure}
    \centering
    \plotone{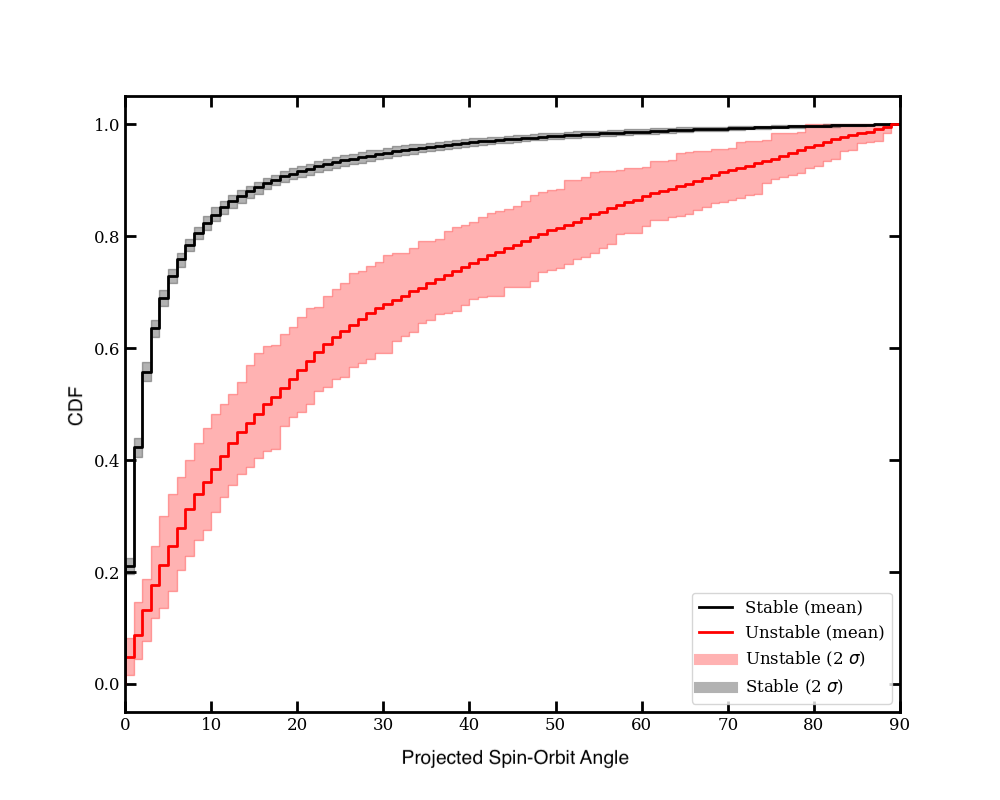}
    \caption{Cumulative distributions of the average projected spin-orbit angle after 10 Myrs for stable and unstable systems. Our Small\_Fixed, Medium\_Fixed, and Large\_Fixed simulations are co-added. The shaded regions denote the $5^{th}$ and $95^{th}$ percentiles of projected spin-orbit angle for each bin.}
    \label{fig:ks_hist}
\end{figure}

If we assume that the planets in our systems begin in an aligned configuration with the stellar spin axis, we can use their orbital inclination as a proxy for the spin-orbit angle $\Psi$. We must then consider how the signature of misalignment would be maintained along a hypothetical line of sight. To ensure that the distribution of planetary spin-orbit angles for stable and unstable populations remain distinct, we choose a random observing angle for each system and calculate the projected spin-orbit angle $\lambda$. We then can utilize the Kolmogorov-Smirnov test to conclude if we can discard the null hypothesis that the \textit{projected} spin-orbit angles of our stable and unstable systems are drawn from the same parent distribution. We repeat this process of choosing a random viewing angle for each system in the population 100 times, and in all 100 cases we obtain a p-value from our KS test that is less than $1\times 10^{-3}$. Therefore, even when projecting our systems along a hypothetical viewing angle, the distributions remain statistically distinct between stable and unstable systems. A cumulative distribution of this test for our fixed binary eccentricity simulations is shown in Figure \ref{fig:ks_hist}, where the median value of the projected angle is plotted for the stable and unstable systems and the shaded regions denote the $5^{th}$ and $95^{th}$ percentiles. This N-body gravitational interaction pathway to misalignment is significant, in our simulations, and bolsters the observations that high obliquity planetary systems likely arise from planet scattering events \citep{win10}. Roughly 1/3 of our systems that experienced instability attain projected spin-orbit angles over 30$^{\circ}$, while only $\sim$1/20 of our stable systems do so.

While our highest eccentricity systems are very likely to be 1-planet systems, this is not the case for our highest inclination planets that have survived instabilities. Of the 16 unstable planetary systems driven to spin-orbit angles over 90$^{\circ}$, 12 of them contain both Jupiter and Saturn analogs. In addition, the Jupiter and Saturn analogs in our 12 multi-planet systems all have final inclinations within $\sim$5$^{\circ}$ of one another (as well as modest eccentricities of $\lesssim0.1$), indicating that they have attained their high inclinations via rigid disk precession \citep{ina97}. 

Also in contrast to planetary eccentricities, there is some correlation between spin-orbit angle and the retention of a binary companion. If we co-add all of our unstable systems and examine those that lost their binary, the median predicted spin-orbit angle is 11.1$^{\circ}$. Meanwhile, for those unstable systems that retain their binary, this median angle is over twice as large at 23.5$^{\circ}$. This suggests that the retention of the binary leads to continued evolution of planetary inclinations (in particular, rigid disk precession) after the instability.

\begin{figure}
    \centering
    \plotone{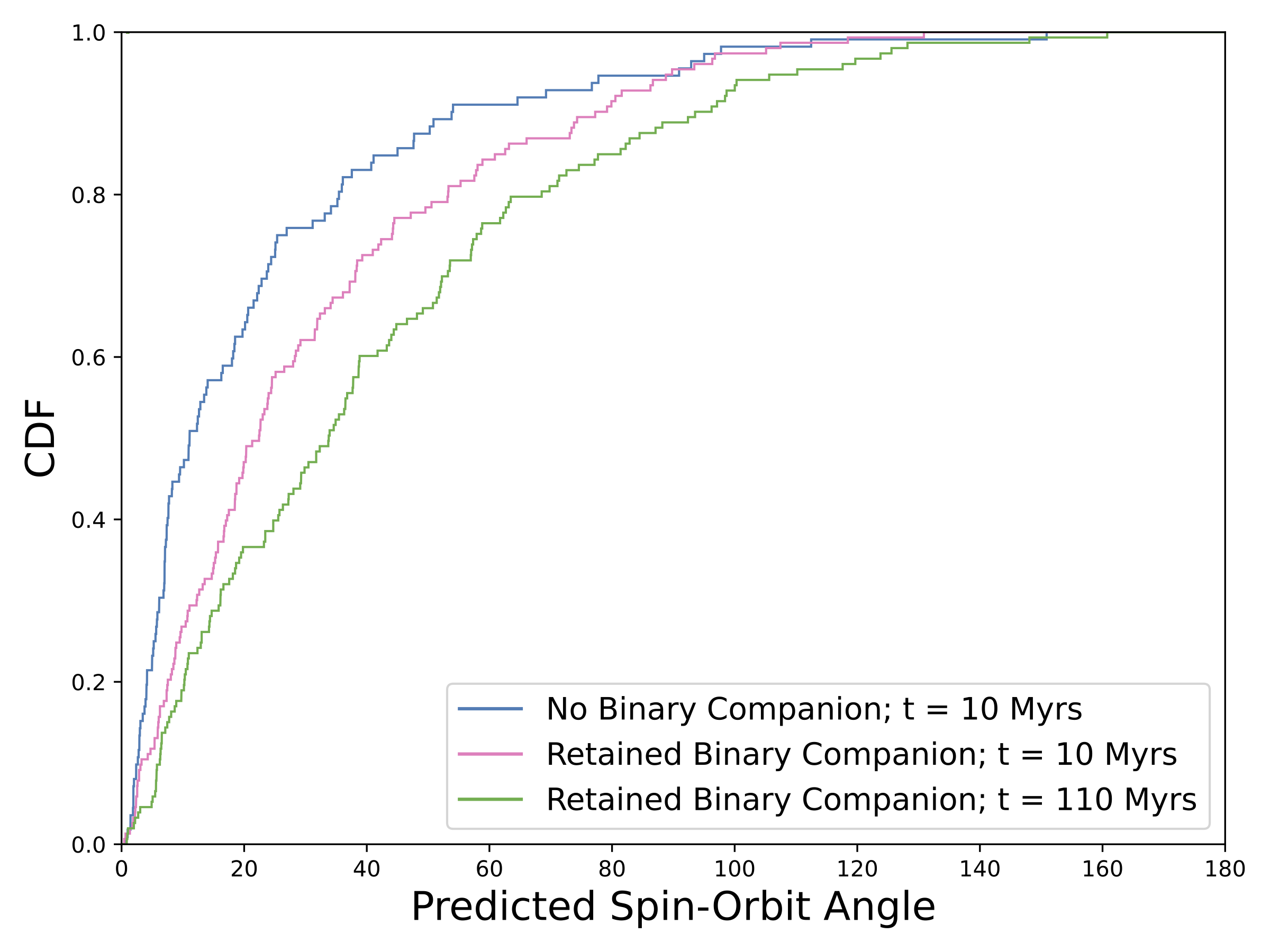}
    \caption{Cumulative distributions of the predicted spin-orbit angle for unstable systems (co-added across all simulation batches). Shown are unstable systems that have lost their binary companion ({\it blue}) and unstable systems that have retained their binary companion after 10 Myrs of integration ({\it pink}) and 110 Myrs of integration ({\it green}).}
    \label{fig:IncBinCorr}
\end{figure}

Given the potentially long timescales for rigid disk precession (see Figure \ref{fig:stepbinary} for instance), this raises the prospect that amongst our unstable systems with a binary companion, the spin-orbit inclinations may continue to evolve after our 10-Myr integration times. To test this possibility, we perform follow-on integrations of our unstable systems that possess a binary for another 100 Myrs. The results of these integrations are shown in Figure \ref{fig:IncBinCorr}, where we plot their distribution of spin-orbit angles. Here we see that systems with binary companions continue to diverge from those that lost their companions. After the additional 100 Myrs of integration, the median spin-orbit angle has increased from 23.5$^{\circ}$ to 33.8$^{\circ}$. In addition, the fraction of systems with retrograde angles has nearly doubled from 6.5\% to 11.1\%. (It should be noted that our distribution of spin-orbit angles after 50 Myrs of integration does not exhibit any statistical differences from our 100-Myr sample, suggesting this suite of systems has reached a dynamical equilibrium by the end of its inegration.) 

This small but significant population of retrograde systems possesses distinct properties. Mostly having achieved their spin-orbit angles via rigid disk precession, the majority (70\%) are multi-planet systems with modest eccentricities ($\bar{e} \simeq 0.09$). Such a set of observed systems would be very consistent with a dynamical history where a planetary system is perturbed by a retained binary companion whose orbit is modified within a birth cluster. 

\subsection{Ramifications for Exoplanets of Single Stars}
\label{singletons}

As mentioned previously, planetary instabilities in our simulations are often accompanied by the dissociation of the binary companion because both processes are tied to strong cluster perturbations on the binary orbit. In fact, $\sim$41\% of our systems that had their binary companion stripped also underwent a planetary orbital instability. Meanwhile, it has been argued that most stars form with one or more stellar companions, but lose their companion via dynamical processing (stellar encounters and tidal stripping) within their birth cluster \citep[e.g.][]{kro95}. Although only a relatively small portion of our simulated systems undergo binary-triggered planetary instabilities (the highest instability rate reached in our simulation sets is $\sim$15\%), our simulations do not encompass all parameter space of binaries or cluster environments. Moreover, our simulations do not consider substructure within our cluster potentials and make other idealized assumptions such as spherical symmetry and instantaneous dissipation. If the actual physical environments of clusters facilitate higher binary dissociation rates and these processes are responsible for a large fraction of the isolated stars observed in the field, a significant fraction (tens of percent) of planetary systems around isolated stars may be subject to instabilities analogous to those in the simulations we present here.

\subsection{The Importance of Gas Potential}

The bulk of the mass in our cluster environments is in the form of a Plummer potential meant to represent the gaseous component. On the other hand, Figure \ref{fig:close_hist} demonstrates that encounters with other cluster stars are critical in many of the planetary system instabilities that our simulated binaries trigger, so one may question whether the gas potential is important to the dynamics of this process. To assess this, we perform two simulation sets (NonStellar\_Fixed and NonStellar\_Wider) that are repeats of Medium\_Fixed and Medium\_Wider. Only this time, they are performed in the NonStellar cluster environment in which stellar masses (but not the Plummer mass) are reduced by an order of magnitude. 

In Table \ref{tab:instabsum}, we see that the rate of planetary instabilities is indeed lower in both reduced stellar mass runs, but the effect is more severe if we use a binary eccentricity distribution fixed at 0.5. Here we find that the instability rate drops by over 75\% (from 9\% of all systems to 2\%). In contrast, when we employ wider binaries with a thermalized eccentricity distribution there is a more modest decrease in instability rate of 44\% (from 14.4\% of all systems to 8\%). The reason for this more modest change is that our wider binary orbits are more easily perturbed and many more of them are initialized with periastra near values necessary to destabilize our planetary systems. Thus, more systems do not require powerful stellar encounters to dramatically drive peristra low enough for a planetary instability, and this can be accomplished with more modest gaseous perturbations. Thus, the importance of the gas potential increases as one considers more eccentric and more weakly bound binaries.

\subsection{Comparison with Past Work}
\subsubsection{System Stability}

As discussed in our introduction, there is an extensive literature studying the stability of planets within binaries and star clusters. Many of the works focused on the stability of planetary systems within clusters only consider the direct effects of stellar flybys, and this limits the rate of planetary instabilities. For instance, \citet{ada06} consider the rate that stellar flybys perturb solar system architectures within embedded clusters of 100--1000 stars, and they find only $\sim$1\% of Neptune-like planets around Sun-like stars are destabilized (their Tables 3 and 7). Similarly, \citet{pro09} consider an even broader range of cluster parameters and typically find destabilization rates of 0.1--1\% for Neptune-like planets (their Table 8). Meanwhile, we show here that the addition of a wide binary companion boosts these destabilization rates by an order of magnitude within comparable cluster environments. To achieve our binary-induced destabilization rates around singleton stars requires much more extreme cluster environments. For instance, \citet{mal11} find that typical open cluster environments yield a destabilization probability of 5--15\% for solar system-like architectures, but their cluster stellar densities are 1--2 orders of magnitude greater than ours, and their cluster lifetimes are 20 times as long. Similarly, \citet{hao13} find cluster-generated stellar flybys can disrupt 99\% of solar system-like architectures, but to achieve this rate requires cluster densities $\sim$2 orders of magnitude greater than ours and 100-Myr lifetimes. If we examine their disruption rates at 10 Myrs (their Figure 3), it is only $\sim$2--10\% with, again, a much denser cluster than any of ours. Our simulations show that a wide binary companion acts as a potential conduit allowing perturbations from the cluster environment to greatly boost the rate of planetary instabilities.

Among works studying planetary stability within binaries, many do not consider temporal evolution of the binary orbit. \citet{hol99} developed a widely employed empirical formula for the maximum semimajor axis, $a_c$, of stable 1-planet systems in the presence of a coplanar wide binary companion with a given mass and eccentricity:

\begin{multline}
a_c = [(0.464 \pm 0.006) + (-0.380 \pm 0.010)\mu + (-0.631 \pm 0.034)e \\
+ (0.586 \pm 0.061)\mu e + (0.150 \pm 0.041)e^2 + (-0.198 \pm 0.074)\mu e^2]a_b
\label{Eq:crita}
\end{multline}
where $a_b$ and $e$ are the binary companion's semimajor axis and eccentricity, and $\mu$ is the ratio of the companion mass to the system mass. This formula is approximate \citep[e.g.][]{qua18}, but if we take Neptune's semimajor axis of 30 au and note our fixed binary mass ratio of 1/3, it allows us to invert Equation 1 and predict a ``critical binary eccentricity'' above which our planetary system should be unstable and below which it will be stable:

\begin{equation}
e_c = \frac{.436-\sqrt{0.19-0.336(0.337 - \frac{30}{a_b})}}{0.168}
\label{Eq:crite}
\end{equation}

If we then examine our simulated initial binary eccentricities, we find that the vast majority (97.1\%) are started on orbits that will allow Neptune to remain stable according to Equation \ref{Eq:crite}. When we examine the 2.9\% of binaries whose initial eccentricities exceed $e_c$ we find that only 1/3 (or less than 1\% of our total systems) end up experiencing a planetary instability, underscoring the fact that Equations \ref{Eq:crita} and \ref{Eq:crite} are approximate estimates of stability. Thus, without considering cluster-driven binary evolution, a stability analysis of our systems would predict nearly all of our systems to be stable (hence their stability over an isolated integration). However, our simulations record 395 planetary instabilities (9.6\% of our integrated systems), and 90\% of these unstable systems begin with binary eccentricities below $e_c$. Of course, our Medium\_Uniform simulation is our only batch of systems with an unskewed binary eccentricity distribution, but if we examine only these systems, we find that 88\% of our unstable systems begin below $e_c$, and their median initial value is 0.83$e_c$. 

Nearly half of our unstable systems lose their binary companion, but for those that retain it, we can study how the final binary eccentricities compare to $e_c$. In these systems, we find that the large majority ($\sim$80\%) finish with binary eccentricities {\it above} $e_c$. This demonstrates that it is relatively rare in our simulations for binaries to evolve to a destabilizing periastron and then evolve to a more modest periastron after triggering an instability. Our cluster environments tend to disperse before this can occur. Hence, final binary eccentricities relative to Equation 2 are a relatively accurate marker of what types of planets could have been destabilized during the cluster phase. However, these binary orbits should not be taken to conclude that such planets could never have existed early in systems' histories. 

\subsubsection{Binary-driven Kozai Cycles}

Past work modeling the evolution of binary stars within clusters has noted that cluster perturbations can excite the orbital inclinations of binary stars and initiate Kozai cycles within planetary systems, possibly driving instabilities and planetary eccentricity excitation. \citet{par09kozai} find that $\sim$20\% of binaries can attain inclinations high enough ($i\gtrsim40^{\circ}$) to excite planetary Kozai cycles. Our simulations find a comparable fraction; 14.8\% of our binary orbits are excited beyond to $i>40^{\circ}$. However, this is not a guarantee of an instability within our planetary systems. Of those systems whose binary orbit is excited beyond $i\gtrsim40^{\circ}$, 60\% did not experience a planetary instability. This is because Kozai cycles tend to be suppressed in multi-planet systems, and they instead undergo rigid disk precession without substantial eccentricity excitation \citep{ina97}. Moreover, of our systems that undergo instability and retain a binary, the final binary inclination is below 40$^{\circ}$ for 60\% of systems. This underscores that Kozai cycles are not the primary driver of instability and eccentricity excitation in our multi-planet systems. Instead, it is the periastron evolution of the binary under cluster perturbations (which is sometimes accompanied by significant binary inclination evolution). 
 
\section{Conclusions}
\label{conclude}

The simulations we present here demonstrate that wide (100 $\gtrsim a_b \gtrsim$ 1000 au) binary stars likely have an early phase with active temporal orbital evolution while they still inhabit their birth clusters. This binary orbital evolution can in turn destabilize the planetary systems they host, touching off episodes of planet-planet scattering. The instabilities we document are typically triggered as cluster perturbations drive the binary periastron toward the planets, exciting one or more planets onto crossing orbits. Close encounters with other cluster stars are generally the cluster perturbation that modifies the binary orbit, but the tidal potential of cluster gas becomes increasingly relevant as we consider more eccentric and/or more weakly bound binaries. Our simulations display a modest rate of instabilities ranging from 5--15\% of all planetary systems. However, due to computational limitations, many regimes of the potential parameter space of embedded star clusters are unexplored. It is possible that other cluster environments as well as different binary and planetary system properties could yield substantially higher rates of planetary instability. 

The planetary instabilities that the binaries trigger often cause significant alterations of the planetary architectures. Namely, the lower mass planets (Uranus and Neptune analogs in our case) are preferentially lost, while the surviving planets (typically Jupiter and Saturn analogs) remain with more excited eccentricities. In addition, even though our simulations begin with coplanar binaries, the inclinations of our planets are often altered as well. This is because the same perturbations that alter binary periastra that trigger instabilities also alter binary inclinations. These heightened binary inclinations cause the planetary systems to undergo rigid disk precession \citep{ina97}, which can ultimately lead to high, even retrograde, spin-orbit angles. Notably, it is rigid disk precession, rather than Kozai cycles, that most often generate simulated systems with the highest spin-orbit angles.

In systems with the highest spin-orbit angles, the binary companion that torques the planets is usually retained, but overall, roughly half of our unstable planetary systems also lose their binary, as the cluster perturbations driving binary orbital evolution can lead to binary unbinding. This raises the prospect that some exoplanets with high eccentricities and/or spin-orbit angles around isolated field stars may be the products of cluster-driven evolution of early binary companions that are lost early in the systems' histories. 

\section{Data Availability Statement}

The simulation data underlying this article are available on the Harvard Dataverse public repository.

\section{Acknowledgements}

This work was performed with support from NASA Exoplanets Research Program grant 80NSSC19K0445, NSF grant AST-1814762, and NSF CAREER Award 1846388. Our computing was performed at the OU Supercomputing Center for Education \& Research (OSCER) at the University of Oklahoma (OU).

\bibliographystyle{aasjournal.bst}
\bibliography{bibliography.bib}
\end{document}